\documentclass[conference]{IEEEtran}
\IEEEoverridecommandlockouts

\usepackage{amsmath,amssymb,amsfonts}
\usepackage{algorithmic}
\usepackage{graphicx}
\usepackage{textcomp}
\usepackage{xcolor}

\usepackage{algorithmic}  
\usepackage{pifont}  
\usepackage{bbding}   
\usepackage{latexsym} 
\usepackage{booktabs}

\usepackage{enumitem}
\usepackage{wrapfig}
\usepackage{color}    
\usepackage{subfigure}    
\usepackage{framed}   
\usepackage{bigstrut}
\usepackage{multirow}

\usepackage[utf8]{inputenc}
\usepackage{newunicodechar}
\usepackage{pdfpages} 

\usepackage{cite}
\usepackage{hyperref} 

\hypersetup{
}

\def\BibTeX{{\rm B\kern-.05em{\sc i\kern-.025em b}\kern-.08em
    T\kern-.1667em\lower.7ex\hbox{E}\kern-.125emX}}

\makeatletter
\newcommand{\linebreakand}{%
  \end{@IEEEauthorhalign}
  \hfill\mbox{}\par
  \mbox{}\hfill\begin{@IEEEauthorhalign}
}
\makeatother

\begin{document}
\title{
Automated Radiology Report Generation Based on Topic-Keyword Semantic Guidance
\thanks{\IEEEauthorrefmark{2}~Corresponding author.}
\thanks{\IEEEauthorrefmark{1}~Jing Xiao and Hongfei Liu contributed equally. This work is supported by the National Nature Science Foundation of China (Project No. 62177015).}
}

\author{
  \IEEEauthorblockN{Jing Xiao\IEEEauthorrefmark{1}\IEEEauthorrefmark{2}}
  \IEEEauthorblockA{
    \textit{School of Computer Science} \\
    \textit{South China Normal University} \\
    Guangzhou, China \\
    xiaojing@scnu.edu.cn
  }
  \and
  \IEEEauthorblockN{Hongfei Liu\IEEEauthorrefmark{1}}
  \IEEEauthorblockA{
    \textit{School of Artificial Intelligence} \\
    \textit{South China Normal University} \\
    Foshan, China \\
    2023025174@m.scnu.edu.cn
  }
  \and
  \IEEEauthorblockN{Ruiqi Dong}
  \IEEEauthorblockA{
    \textit{School of Artificial Intelligence} \\
    \textit{South China Normal University} \\
    Foshan, China \\
    2023025193@m.scnu.edu.cn
  }
 \linebreakand
  \IEEEauthorblockN{Jimin Liu}
  \IEEEauthorblockA{
    \textit{Department of Biomedical Engineering} \\
    \textit{National University of Singapore} \\
    Singapore \\
    liujm@nus.edu.sg
  }
  \and
  \IEEEauthorblockN{Haoyong Yu}
  \IEEEauthorblockA{
    \textit{Department of Biomedical Engineering} \\
    \textit{National University of Singapore} \\
    Singapore \\
    bieyhy@nus.edu.sg
  }
}

\maketitle

\begin{abstract}
Automated radiology report generation is essential in clinical practice. However, diagnosing radiological images typically requires physicians 5-10 minutes, resulting in a waste of valuable healthcare resources. Existing studies have not fully leveraged knowledge from historical radiology reports, lacking sufficient and accurate prior information. To address this, we propose a Topic-Keyword Semantic Guidance (TKSG) framework. This framework uses BiomedCLIP to accurately retrieve historical similar cases. Supported by multimodal, TKSG accurately detects topic words (disease classifications) and keywords (common symptoms) in diagnoses. The probabilities of topic terms are aggregated into a topic vector, serving as global information to guide the entire decoding process. Additionally, a semantic-guided attention module is designed to refine local decoding with keyword content, ensuring report accuracy and relevance. Experimental results show that our model achieves excellent performance on both IU X-Ray and MIMIC-CXR datasets. The code is available at \url{https://github.com/SCNU203/TKSG}
\end{abstract}


\section{Introduction}

\begin{figure}
    \centering
    \includegraphics[width=\columnwidth]{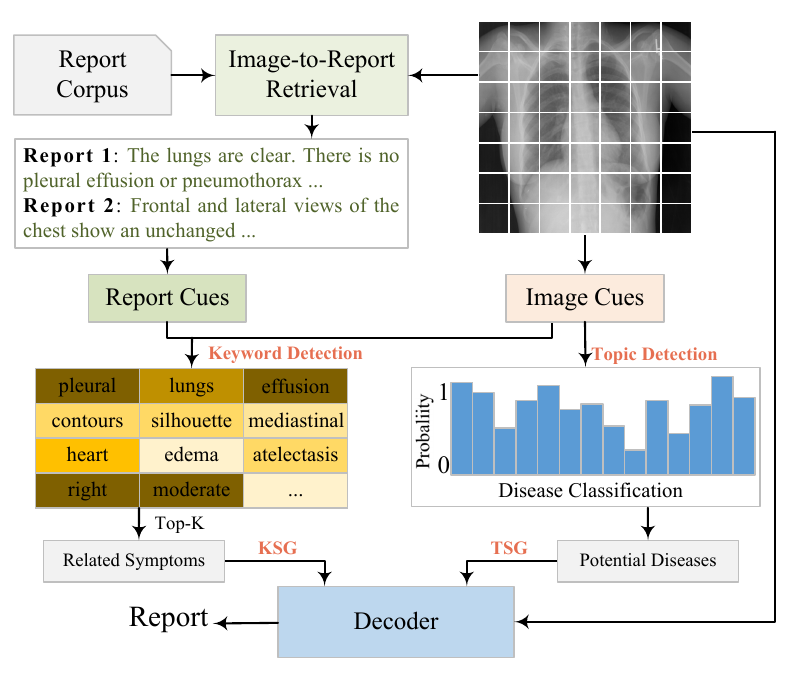} 
    \caption{Diagram of the Topic-Keyword Semantic Guidance (TKSG) framework. TKSG aims to provide accurate and comprehensive topic and keyword information for report generation.}
    \label{fig:overview}
    \vspace{-15pt}
\end{figure}

Radiological images, such as X-rays and MRIs, are crucial for disease diagnosis, but traditional report generation~\cite{ganeshan2018structured} is often time-consuming and error-prone. Advances in computer vision and large datasets have made automated radiology report generation~\cite{zhang2023knowledge} an appealing alternative, reducing radiologists' workload and improving diagnostic efficiency.

\begin{figure*}[t]
    \centering
    \includegraphics[width=\textwidth]{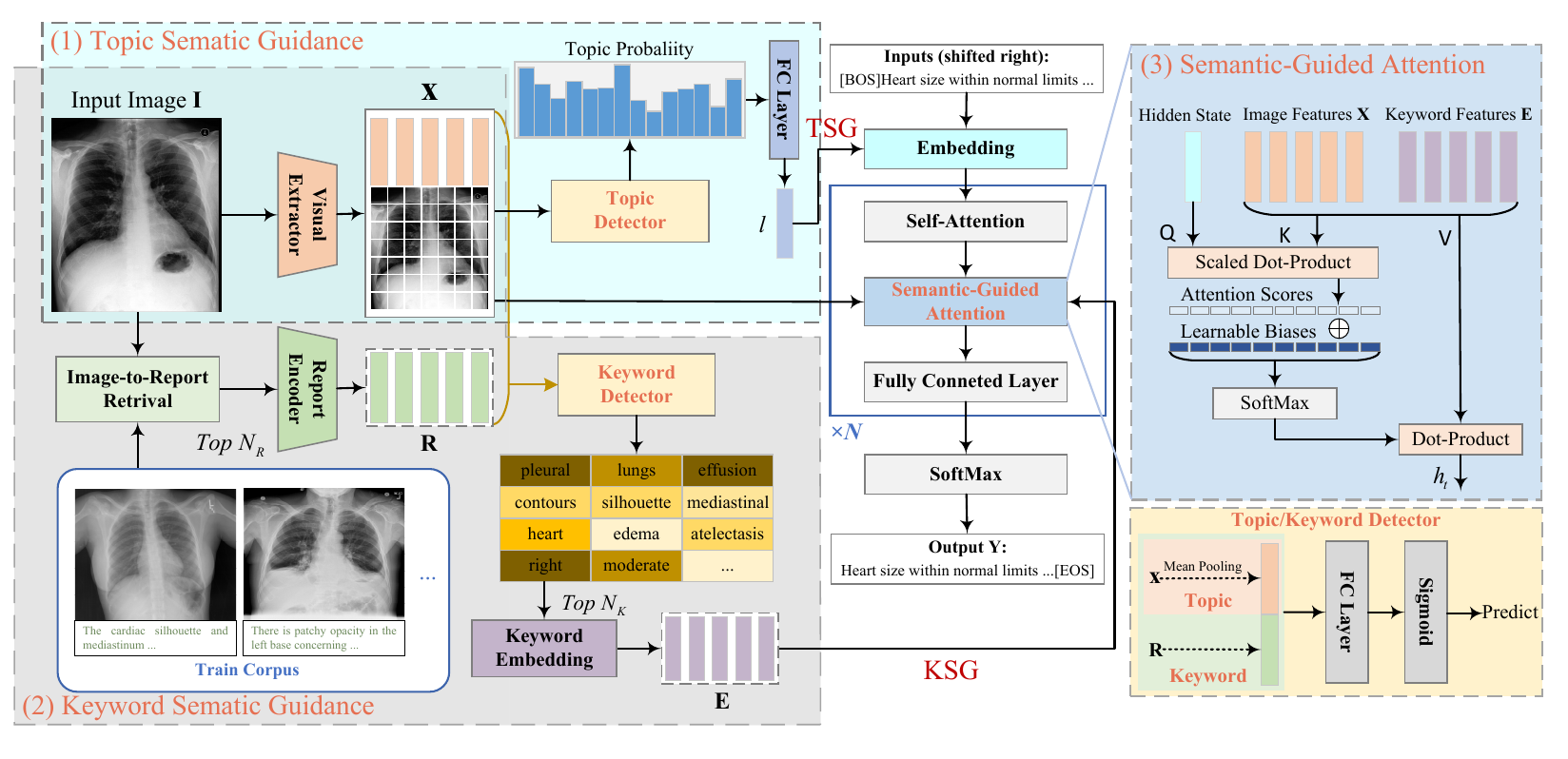} 
    \caption{Overview of the proposed TKSG. 1) Topic Sematic Guidance (TSG): This module predicts the probability of 14 diseases based on visual features and aggregates these probabilities into topic vectors $l$ for global semantic guidance (Section \ref{sec:topic_detector}). 2) Keyword Sematic Guidance (KSG): This module combines the multimodal information, predicts the probabilities of concepts and extracts the top $N_K$ concept as keywords for further refinement analysis (Section \ref{sec:keyword_detector}). 3) Semantic-Guided Attention Module: Fusion of visual and keyword features for local semantic guidance (Section \ref{sec:report_generator}).}
    \label{fig:model}
   \vspace{-15pt}
\end{figure*}

Existing methods~\cite{R2Gen,R2GenCMN,M2KT} have improved report generation but still struggle with producing precise free-text reports due to insufficient utilization of historical reports and visual information. To address this, we propose a model that first assesses potential diseases globally through topic analysis and then provides detailed symptom descriptions locally using keyword analysis. For instance, the model may identify "\textit{right lung atelectasis}" (topic information) from image features and "\textit{moderate right pleural effusion}" (keyword information) from images and similar historical case, generating a report like "\textit{moderate right pleural effusion with right lung atelectasis}."

To achieve this vision, we first use CheXbert~\cite{CheXbert}, an automatic labeling tool, to generate the pseudo-labels of disease classification for each report. Next, we leverage image information to predict the classification of these diseases, thereby forming topic vectors, serving as Topic Semantic Guidance (\textbf{TSG}). This process helps the model understand the condition from a global perspective and provides more accurate guidance for subsequent local analysis, such as detailed symptom. 

Then we explore image-text association methods in medical models proposed in recent years, such as BiomedCLIP~\cite{biomedclip} and MedCLIP~\cite{medclip}. With these models, we retrieve the most similar historical case reports and utilize multimodal fusion information to predict keywords that provide more detailed symptom descriptions for these diseases, serving as Keyword Semantic Guidance (\textbf{KSG}).

Building on the above idea, we propose an innovative Topic-Keyword Semantic Guidance (TKSG) framework, with its report generation process illustrated in  Fig~\ref{fig:overview}, which first captures the global context of a condition through topic terms and then meticulously analyzes specific causes at a local level through keyword refinement. 
This approach promises to better integrate historical data and visual information, enhancing the precision of report generation. Our main contributions are summarized as follows:

\begin{itemize}[noitemsep,label=\textbullet]
    \item We propose the TKSG model, which predicts topic terms and keywords through multimodal information, guiding global report generation and enriching prior knowledge with semantic-guided attention and keywords.
     \item Through a large number of experiments, we compare the impact of using BiomedCLIP, MedCLIP, and CLIP~\cite{clip} models to retrieve reference reports on the performance of report generation, effectively realizing the knowledge transfer in radiological report generation.
     \item Experimental results show that our model achieves state-of-the-art performance in almost all evaluation metrics on the public datasets IU-Xray~\cite{IU} and MIMIC-CXR~\cite{MIMIC}.
\end{itemize}

\section{Related Work}

\vspace{-0.3em}
\textbf{Radiology Report Generation.} 
Radiology report generation is closely related to image captioning~\cite{MultiCapCLIP}, aiming to describe image content using sentences. Some works~\cite{miura2020improving} focused on generating reports with clinical accuracy, employing reinforcement learning to improve fidelity.
There are also studies proposing template-based approaches~\cite{kale2023replace} to improve the quality and consistency of reports. 
Meanwhile, recent research has focused on multimodal alignment and learned knowledge bases, aiming to integrate information from diverse modalities to improve cross-modal task performance~\cite{M2KT}.


\textbf{Cross-Modal Learning in Medical Field.} 
In the medical field, cross-modal learning has shown excellent application potential in medical image-text matching. 
Transformer architectures play a crucial role in addressing fine-grained differences between images and texts, such as the memory-driven R2Gen~\cite{R2Gen} and the progressive M2TR~\cite{M2TR} models.
Cross-modal retrieval techniques efficiently match images and texts. 
The FSE model proposed by Liu et al~\cite{FSE}. learns unimodal features through cross-modal alignment between images and factual serialization in the report.

\section{Methodology}

In the radioactivity report generation task, the input source to the model is $\mathbf{X}=\{ \mathbf{x_1},\mathbf{x_2}, \ldots, \mathbf{x_n}, \ldots, \mathbf{x_N} \}$, where $\mathbf{x_n} \in \mathbb{R}^{d_h}$, denoting the result extracted from the original radioactivity image $\mathbf{I} \in \mathbb{R}^{H \times W \times C}$ by a visual feature extractor, where $H$, $W$ and $C$ are the height, width and the number of channels of an image. Corresponding to the original image $\mathbf{I}$ is the target radiological report $\mathbf{Y}=\{ y_1,y_2,\ldots,y_t,\ldots,y_T \}$, where $y_t \in |V|$, denoting the generated tokens; $T$ is the length of the report, and $|V|$ is the size of the vocabulary. Thus, the complete flow of the entire report generation can be formalized as:
\vspace{-0.3em}
\begin{equation}
 p(\mathbf{Y}|\mathbf{I})=\prod^{T}_{t=1}p(y_t|y_1,\ldots,y_{t-1},\mathbf{I})
\end{equation}

\vspace{-0.3em}
The goal of training is to minimize the loss function $\mathcal{L}_{rep}$:
\vspace{-0.3em}
\begin{equation}
\mathcal{L}_{rep}=-\frac{1}{T} \sum^{T}_{t=1}{l(y_t) \cdot \log p({y^*_t|\mathbf{Y}^*_{<t},\mathbf{I}})}
\end{equation}

\vspace{-0.3em}
where $l(y_t)$ is the one-hot vector of the target word, and $p({y^*_t|\mathbf{Y}^*_{<t},\mathbf{I}})$ is the probability distribution of the word predicted by the model based on the image features $\mathbf{I}$ and previous predictions.

Figure~\ref{fig:model} shows our proposed TKSG model, with details provided in the following subsections.
\subsection{Visual Extractor}
\label{sec:ve}
Given a radiology image $\mathbf{I}$, an image processor is first used to split $\mathbf{I}$ into $N$ patches. Then we use the Swin Transformer Tiny~\cite{swin}, denoted as $f_{ve}(\cdot)$, to extract the visual features of the image. 
Finally, a linear layer is applied to map the visual features to a $d_h$-dimension space to get the final visual representations $\mathbf{X}$:
\vspace{-0.3em}
\begin{equation}
 \mathbf{X}= LN(f_{ve}(\mathbf{I}) \mathbf{W}^{enc} + b^{enc}) =\{ \mathbf{x_1},\mathbf{x_2}, \ldots, \mathbf{x_N} \}
\end{equation}

\vspace{-0.3em}
where $LN$ denotes the normalization layer, where $\mathbf{W}^{enc} \in \mathbb{R}^{d_b \times d_h}$ and $b^{enc} \in \mathbb{R}^{d_h}$ are trainable parameters and $d_b$ denotes is $f_{ve}(\cdot)$ output dimension.
\subsection{Image-to-Report Retrieval}
\label{sec:re}

We perform image-to-report retrieval using BiomedCLIP~\cite{biomedclip}, MedCLIP~\cite{medclip}, and CLIP~\cite{clip}, and conduct experiments to explore their impact on our model’s performance. Specifically, the text encoders in these models are decoder-only Transformers. Given a set of candidate reports $\mathbf{Z}$, we extract the activation values of the top [EOS] tokens as report features. For a query image, we use an image encoder to obtain mean-pooled features. By calculating the cosine similarity between the image and report features, we generate a similarity vector and select the top $N_R$ most relevant reports.

\subsection{Topic Detector}
\label{sec:topic_detector}

In order to identify pseudo disease labels that may be associated with certain diseases in the input images, we use the CheXbert~\cite{CheXbert} model. This is a labeling generation tool that automatically generates radiology reports and is capable of extracting $N_T=14$ different disease labels from the reports. 
With this model, we generate the corresponding pseudo disease labels for each image. We define topic labels as $\mathbf{T}$:
\vspace{-0.3em}
\begin{equation}
 \mathbf{T} = f_{cx}(\mathbf{Y}) = \{t_1,t_2,t_3,\ldots,t_{14}\} 
\end{equation}

\vspace{-0.3em}
Here, $f_{cx}(\cdot)$ denotes the automatic radiology report labeler. The result is represented as a one-hot vector where $t_i\in \{0,1\}$. If the topic is present in the report, $t_i=1$; otherwise, $t_i=0$.

Next, we perform mean pooling on the image features and input them into Topic Detector to generate the patient's topic probabilities $P_T$ (Eq.~\ref{eqn:td}), assisting the model in identifying potential diseases. Then, the topic probabilities $P_T$ are aggregated into topic vector $l \in \mathbb{R}^{d_h} $ using a FCLayer (Eq.~\ref{eqn:fc}).
\vspace{-0.5em}
\begin{equation}
\label{eqn:td}
 P_T = Topic\_Detector(x) = Sigmoid(x\mathbf{W}^{TD}+b^{TD}) 
\end{equation}
\begin{equation}
\label{eqn:fc}
 l = FCLayer(P_T) 
\end{equation}

\vspace{-0.5em}
where $p_i \in P_T$ denotes the predicted probability of the $i$ topic, while $\mathbf{W}^{TD} \in \mathbb{R}^{d_h \times 14}$ and $b^{TD} \in \mathbb{R}^{14}$ are trainable parameters. We define $x=f(\mathbf{X}) \in \mathbb{R}^{d_h}$ and $f{(*)}$ denotes mean pooling.

We train Topic Detector with the following loss function:
\vspace{-0.3em}
\begin{equation}
 \mathcal{L}_{td}=-\frac{1}{N_T}\sum^{N_T}_{i=1}(t_i\log{p_i}+(1-t_i)\log{(1-p_i)}) 
\end{equation}

\subsection{Keyword Detector}
\label{sec:keyword_detector}

Before entering the Keyword Detector module, we first select the $N_W$ most frequently used words (e.g. nouns, verbs, and adjectives) from the report as concepts, while removing stop words. Then, for each image, we generate a multi-hot keyword label: $C = \{c_1, c_2, \ldots, c_{N_W}\}$. If the report corresponding to the image contains the $i^{th}$ keyword, then $c_i = 1$; otherwise, $c_i = 0$.

Next, we define the input to the Keyword Detector as $x$, and the process can be further detailed in Eq.~\ref{eqn:kd} as follows:
{\small
\begin{equation}
\label{eqn:kd}
 P_W = Keyword\_Detector(x) = Sigmoid(x\mathbf{W}^{KD}+b^{KD}) 
\end{equation}}

where $p_i \in P_W$ represent the predicted probability of the $i^{th}$ concept. The parameters $\mathbf{W}^{KD} \in \mathbb{R}^{2d_h \times N_W}$, $b^{KD} \in \mathbb{R}^{N_W}$ are trainable. We define $x = [f(\mathbf{X}); f(\mathbf{R})] \in \mathbb{R}^{2d_h}$, where $f(*)$ denotes mean pooling, $[;]$ denotes concatenation, and $\mathbf{R}$ denotes features of similar historical case reports.
We train Keyword Detector with the following loss function:
\vspace{-0.3em}
\begin{equation}
 \mathcal{L}_{kd} = -\frac{1}{{N_W}}\sum^{N_W}_{i=1}(c_i\log{p_i}+(1-c_i)\log{(1-p_i)}) 
\end{equation}


\vspace{-0.3em}
Next, we select the top $N_K$ words with the highest probability from concepts,
constituting the set of keywords $\mathbf{S}=\{s_i\}^{N_K}_{i=1}$, where $s_i \in [1,N_W]$, and the features of keywords are denoted as $\mathbf{E}=\{ k_i \}^{N_K}_{i=1} \in \mathbb{R}^{N_K \times d_h}$. 
The derivation of $\mathbf{E}$ is as follows:
\begin{equation}
 \mathbf{E} = KeywordEmb(\mathbf{S}) = LN(\mathbf{W}^{emb}_{[s_i]} + \mathbf{W}^{rank}_{[i]}) 
\end{equation}

where the subscript $[j]$ denotes the $j^{th}$ row of a matrix, $\mathbf{W}^{emb} \in \mathbb{R}^{N_W \times d_h}$ and $\mathbf{W}^{rank} \in \mathbb{R}^{N_K \times d_h}$ denote the trainable matrices for keyword embedding  and rank embedding  respectively. 

\begin{table*}[ht]
  \vspace{-10pt} 
  \caption{TKSG results on the MIMIC-CXR and IU X-Ray test sets. The best results are in \textbf{boldface}, and the \underline{underlined} are the second-best results. *: Results re-implementations using the official code. P and R denote precision and recall, respectively.}
  \vspace{-13pt} 
  \begin{center}
    \begin{tabular}{clcccccccccc}
    \toprule
    \textbf{Dataset} & \multicolumn{1}{c}{\textbf{Model}} & \textbf{Year} & \textbf{BLEU-1} & \textbf{BLEU-2} & \textbf{BLEU-3} & \textbf{BLEU-4} & \textbf{METEOR} & \textbf{ROUGE-L} & \textbf{P} & \textbf{R} & \textbf{F1} \\
    \midrule
    \multicolumn{1}{c}{\multirow{8}[4]{*}{\newline{}\textbf{IU X-ray}}} & R2Gen & EMNLP'20 & 0.470  & 0.304 & 0.219 & 0.165 & -     & 0.371 & -     & -     & - \\
          & R2GenCMN* & ACL'21 & 0.462 & 0.295 & 0.216 & 0.159 & 0.184 & 0.356 & -     & -     & - \\
          & CMM-RL & ACL'22 & 0.494 & 0.321 & \underline{0.235} & \underline{0.181} & 0.201 & 0.384 & -     & -     & - \\
          & M2TR  & MedAI'23 & 0.486 & 0.317 & 0.232 & 0.173 & 0.192 & 0.390  & -     & -     & - \\
          & M2KT  & MedAI'23 & 0.497 & 0.319 & 0.230  & 0.174 & -     & \textbf{0.399} & -     & -     & - \\
          & \multicolumn{1}{p{5.125em}}{FSE} & MedAI'24 & \underline{0.504} & 0.327 & 0.234 & 0.174 & \underline{0.214} & 0.380  & -     & -     & - \\
          & \multicolumn{1}{p{5.125em}}{MA} & \multicolumn{1}{p{5.125em}}{~~AAAI'24} & 0.501 & \underline{0.328} & 0.230  & 0.170  & 0.213 & 0.386 & -     & -     & - \\
\cmidrule{2-12}          & TKSG  & - & \textbf{0.511} & \textbf{0.345} & \textbf{0.247} & \textbf{0.183} & \textbf{0.215} & \underline{0.394} &   -   &    -   &  -  \\
    \midrule
    \multicolumn{1}{c}{\multirow{9}[4]{*}{\textbf{MIMIC-CXR}\newline{}}} & R2Gen & EMNLP'20 & 0.353 & 0.218 & 0.145 & 0.103 & 0.142 & 0.277 & 0.333 & 0.273 & 0.276 \\
          & R2GenCMN* & ACL'21 & 0.354 & 0.213 & 0.138 & 0.099 & 0.137 & 0.271 & \underline{0.449} & 0.339 & 0.387 \\
          & CMM-RL & ACL'22 & 0.381 & 0.232 & 0.155 & 0.109 & 0.151 & \underline{0.287} & 0.342 & 0.294 & 0.292 \\
          & M2TR  & MedAI'23 & 0.378 & 0.232 & 0.154 & 0.107 & 0.145 & 0.272  & 0.240  & \textbf{0.428} & 0.308 \\
          & \multicolumn{1}{p{5.125em}}{M2KT} & MedAI'23 & \underline{0.386} & 0.237 & 0.157 & 0.111 & -     & 0.274  & 0.420  & 0.339 & 0.352 \\
          & \multicolumn{1}{p{5.125em}}{FSE} & MedAI'24 & \textbf{0.396} & \underline{0.241} & \underline{0.162} & \underline{0.116} & \underline{0.154} & 0.279  & -     & -     & - \\
          & \multicolumn{1}{p{5.125em}}{MA} & \multicolumn{1}{p{5.125em}}{~~AAAI'24} & \textbf{0.396} & \textbf{0.244} & \underline{0.162} & 0.115 & 0.151 & 0.274 & 0.411 & 0.398 & \underline{0.389} \\
\cmidrule{2-12}          & TKSG  & - & 0.378 & \underline{0.241} & \textbf{0.165} & \textbf{0.123} & \textbf{0.157} & \textbf{0.297} & \textbf{0.548} & \underline{0.407} & \textbf{0.467} \\
    \bottomrule
    \end{tabular}%
    \vspace{-13pt} 
  \label{tab:contrast}%
  \end{center}
\end{table*}%

\subsection{Report Generator}
\label{sec:report_generator}
\textbf{Topic Semantic Guidance.} At the stage where the model generates report $\mathbf{Y}=\{ y_1,y_2,\ldots,y_t, \ldots, y_T \}$ , each word $y_t$ is generated based on the image feature $\mathbf{X}$ and the previously generated word $\mathbf{Y}_{<t}$. Specifically, at time step $t$, we influence the overall generation process of decoding by fusing the topic vector $l$ into the input embedding of the report generator. 
The formula for calculating the input embedding is as follows:
\begin{equation}
 e_t = Embedding(y_{t-1},l)=LN(\mathbf{W}^{word}_{[y_{t-1}]} + \mathbf{W}^{pos}_{[t]}+l)
\end{equation}

where $\mathbf{W}^{word} \in \mathbb{R}^{|V|\times d_h}$ and $\mathbf{W}^{pos} \in \mathbb{R}^{T_{max} \times d_h}$ denote the trainable matrices for word embedding and positional embedding. $|V|$ is the vocabulary size, and $T_{max}$ is the maximum sequence length of the generated report.

\textbf{Keyword Semantic Guidance.} We concatenate the keyword embeddings $\mathbf{E}$ with the image features $\mathbf{X}$, enabling their integration within the model to provide local semantic guidance. In this way, the keyword embedding $\mathbf{E}$ can provide more precise semantic information for the image features $\mathbf{X}$, so that the model can better focus on and capture the content related to the keywords in the generation phase, as follows: 
\begin{equation}
 h_t = SG\_Att(\mathbf{Q},\mathbf{K},\mathbf{V}) = Softmax(\frac{\mathbf{Q} \mathbf{K}^T}{\sqrt{d_h}}) \mathbf{V} 
\end{equation}

where $\mathbf{Q}$ is the feature projection from the output of the self-attention layer, while $\mathbf{K}$ and $\mathbf{V}$ are the combined features obtained by concatenating the image features and keyword embeddings, denoted as $[\mathbf{X};\mathbf{E}]$.

Finally, we pass the output $h_t$ from the Semantic-Guided Attention module to the classification head to predict the next word $y_t$. The process is as follows:
\begin{equation}
  y_t \sim p(y_t|Y_{<t},V)=Softmax(h_t\mathbf{W}^{cls})
\end{equation}

where $p(y_t|Y_{<t},V) \in \mathbb{R}^{|V|}$ denotes a probability distribution over a vocabulary list $|V|$ and $\mathbf{W}^{cls} \in \mathbb{R}^{d_h \times |V|}$ is a learnable parameter matrix. 

\textbf{Objective Function.}  Our final optimized objective function is defined as $\mathcal{L}_{all}$. 
\begin{equation}
  \mathcal{L}_{all}=\mathcal{L}_{rep}+\mathcal{L}_{kd}+\mathcal{L}_{td}
\end{equation}

\section{Experiments}
\subsection{Datasets}
In our experiments, we use two benchmark datasets: IU X-RAY~\cite{IU} from Indiana University, containing 7,470 chest X-rays and 3,955 reports, and MIMIC-CXR~\cite{MIMIC} from Beth Israel Deaconess Medical Center, the largest public radiographic dataset with 473,057 images and 206,563 reports.

\subsection{Evaluation Metrics and Baselines}
\textbf{NLG.} We use various NLG metrics, including BLEU, METEOR, and ROUGE, to evaluate report quality. These metrics assess the similarity between the generated results and the Ground Truth from different perspectives, ensuring a comprehensive and objective assessment.

\textbf{CE.}\footnote{Note that CE metrics only apply to MIMIC-CXR because CheXbert's labeling schema is specific to MIMIC-CXR, not IU X-RAY.} As NLG metrics insufficiently measure clinical correctness, the CheXbert~\cite{CheXbert} is applied to label the generated reports and compare the results with Ground Truth in 14 different categories related to thoracic diseases. We use precision, recall and F1 to evaluate model performance for CE metrics.


\textbf{Baseline.} To evaluate the performance of TKSG, we compare it with the following state-of-the-art baselines: R2Gen~\cite{R2Gen}, R2GenCMN~\cite{R2GenCMN}, CMM-RL~\cite{CMM-RL}, M2TR~\cite{M2TR}, M2KT~\cite{M2KT}, FSE~\cite{FSE} and MA~\cite{MA}.

\subsection{Implementation Details}
In our experiments, the visual feature extractor employs the Swin Transformer Tiny~\cite{swin} to extract 512-dimensional features from each image. The encoder-decoder architecture is based on a Transformer with 3 layers and 8 attention heads, and a hidden layer dimension of 512, with randomly initialized parameters. We use the ChexBert~\cite{CheXbert} model to extract pseudo-labels and the BiomedCLIP~\cite{biomedclip} model to retrieve similar historical case reports. The number of retrieved reports is set to $N_R=30$, the number of concepts to $N_W=100$, and the number of keywords to $N_K=20$. The model is trained with the Adam optimizer using cross-entropy loss, with initial learning rates of $2 \times e^{-4}$ and $5 \times e^{-4}$, and a decay rate of 0.8 per epoch. During report generation, the beam size is set to 3 to balance effectiveness and efficiency. All experiments are conducted on four NVIDIA 3090 GPUs.

\begin{table*}[ht]
  \vspace{-10pt} 
  \centering
  \caption{Ablation study results on the MIMIC-CXR and IU-Xray datasets. The best values are highlighted in \textbf{boldface}. The average improvement over all NLG metrics compared to BASE is also presented in the "AVG. $\bigtriangleup$" column.}
    \begin{tabular}{cccccccccccc}
    \toprule
    \textbf{Datsets} & \textbf{Model} & \textbf{BLEU-1} & \textbf{BLEU-2} & \textbf{BLEU-3} & \textbf{BLEU-4} & \textbf{METEOR} & \textbf{ROUGE-L} & \textbf{AVG. $\bigtriangleup$} & \textbf{P} & \textbf{R} & \textbf{F1} \\
    \midrule
    \multicolumn{1}{c}{\multirow{4}[2]{*}{\textbf{IU X-ray}}} & BASE  & 0.462 & 0.295 & 0.216 & 0.159  & 0.184 & 0.356 & \textbackslash{} & -     & -     & - \\
          & TSG   & 0.498 & 0.316 & 0.218 & 0.165 & 0.195 & 0.372 & 5.5\% & -     & -     & - \\
          & KSG{\tiny -B} & 0.495 & 0.329 & 0.241 & 0.174 & 0.202 & 0.381  & 9.0\%   & -     & -     & - \\
          & TKSG{\tiny -B} & \textbf{0.511} & \textbf{0.345} & \textbf{0.247} & \textbf{0.183} & \textbf{0.215} & \textbf{0.394} & \textcolor[rgb]{ .2,  .2,  .2}{\textbf{13.4\%}} & -     & -     & - \\
    \midrule
    \multicolumn{1}{c}{\multirow{4}[2]{*}{\textbf{MIMIC-CXR}}} & BASE  & 0.354 & 0.213 & 0.138 & 0.099 & 0.137 & 0.271 & \textbackslash{} & 0.449 & 0.339 & 0.387 \\
          & TSG   & 0.364 & 0.233 & 0.161 & 0.119 & 0.150  & 0.292 & 8.8\% & 0.502 & 0.374 & 0.428  \\
          & KSG{\tiny -B} & 0.371 & 0.235 & 0.162 & 0.118 & 0.147 & 0.287 & 8.9\% & 0.521 & 0.389 & 0.445 \\
          & TKSG{\tiny -B} & \textbf{0.378} & \textbf{0.241} & \textbf{0.165} & \textbf{0.123} & \textbf{0.157} & \textbf{0.297} & \textbf{12.3\%} & \textbf{0.548} & \textbf{0.407} & \textbf{0.467} \\
    \bottomrule
    \end{tabular}%
  \label{tab:ablation}%
  \vspace{-10pt} 
\end{table*}%

\begin{table}[t]
  \small
  \centering
  \caption{Comparison of the best NLG metrics on the IU X-Ray dataset for retrieving similar cases using CLIP, MedCLIP, and BiomedCLIP. "K" denotes keyword guidance, "T" topic guidance, with \checkmark for module usage and \ding{55} for non-usage.}
  \resizebox{\columnwidth}{!}{
    \begin{tabular}{ccccccccc}
    \toprule
    \textbf{Model} & \textbf{K} & \textbf{T} & \textbf{B-1} & \textbf{B-2} & \textbf{B-3} & \textbf{B-4} & \textbf{MTR} & \textbf{RG-L} \\
    \midrule
    \multirow{2}[2]{*}{-} & \ding{55}     & \ding{55}     & 0.462 & 0.295 & 0.216 & 0.159  & 0.184 & 0.356 \\
          & \ding{55}     & \checkmark     & 0.498 & 0.316 & 0.218 & 0.165 & 0.195 & 0.372 \\
    \midrule
    \multirow{2}[2]{*}{CLIP} & \checkmark     & \ding{55}     & 0.490  & 0.321 & 0.227 & 0.168 & 0.193  & 0.374 \\
          & \checkmark     & \checkmark     & 0.500 & 0.338  & 0.243  & 0.178 & 0.209  & 0.380  \\
    \midrule
    \multirow{2}[2]{*}{MedCLIP} & \checkmark     & \ding{55}     & 0.486 & 0.324 & 0.228  & 0.171 & 0.197 & 0.371 \\
          & \checkmark     & \checkmark     & 0.505 & 0.343 & 0.246 & 0.179 & 0.213 & 0.383 \\
    \midrule
    \multicolumn{1}{c}{\multirow{2}[2]{*}{BiomedCLIP}} & \checkmark     & \ding{55}     & 0.495 & 0.329 & 0.231 & 0.176 & 0.202 & 0.381  \\
          & \checkmark     & \checkmark     & \textbf{0.511} & \textbf{0.345} & \textbf{0.247} & \textbf{0.183} & \textbf{0.215} & \textbf{0.394} \\
    \bottomrule
    \end{tabular}}%
    \vspace{-10pt}
  \label{tab:local_global}%
\end{table}%

\subsection{Quantitative Analysis}
\textbf{Comparison with SOTA.} The NLG and CE evaluation results are shown in Table~\ref{tab:contrast}. Our proposed model TKSG achieves the current SOTA performance on both the IU X-ray and MIMIC-CXR datasets, showing significant improvement. In particular, on the IU X-ray dataset, TKSG achieves the best performance across all evaluation metrics except RG-L, including outperforming the second best method by 5.2\% and 5.1\% on the BLEU-2 and BLEU-3 metrics, respectively. On the MIMIC-CXR benchmark dataset, TKSG also performs well, achieving the best performance on all metrics except BLEU-1 and BLEU-2.
Notably, our model outperforms others in clinical accuracy, delivering reports that not only exhibit exceptional language quality but also align more closely with actual diagnoses. This outstanding performance is evident across all CE metrics, except for recall (R), highlighting the superior medical precision and reliability of our approach.

\begin{figure}[t]
    \centering
    \includegraphics[width=\columnwidth]{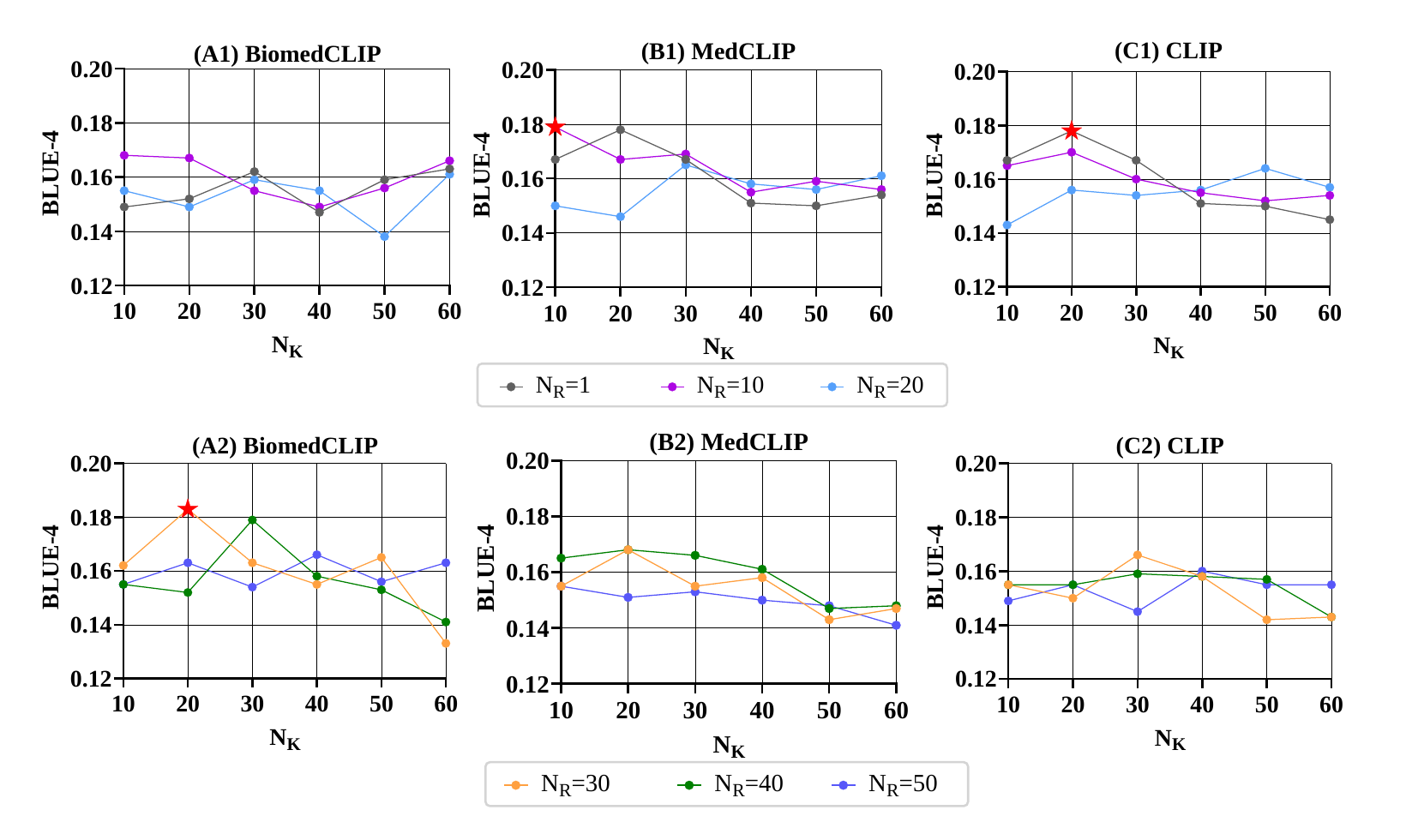} 
    \caption{BiomedCLIP, MedCLIP and CLIP models are applied for topic-keyword semantic guidance analysis on the IU X-ray dataset. \textcolor[rgb]{1,0,0}{\ding{72}}: indicates the best BLUE-4 score.} 
    \label{fig:keyword}
   \vspace{-15pt}
\end{figure}

\textbf{Topic-Keyword Semantic Guidance Analysis.} 
We conduct experiments on the IU X-ray dataset, using BiomedCLIP, MedCLIP, and CLIP models for similar historical case reports retrieval. The results show that our method achieves the best generation performance, as shown in Table~\ref{tab:local_global}, validating the effectiveness of the semantically guided strategy in improving the quality of the generation of reports. Additionally, we further explore the impact of different parameter settings on the semantic guidance effect, as shown in Figure~\ref{fig:keyword}. Specifically, we present the results of 108 tests with the retrieval report count set as $N_R=\{1,10,20,30,40,50\}$, the number of concepts as $N_W=100$, and the number of keywords for local semantic guidance as $N_K=\{10,20,30,40,50,60\}$. The experiments demonstrate that different combinations of parameters significantly affect model generation performance, confirming the key role of optimized parameter selection in improving report generation quality. Ultimately, we select BiomedCLIP as the retrieval model and set the parameters to $N_R=30$, $N_W=100$ and $N_K=20$.

\textbf{Ablation Analysis.} Table~\ref{tab:ablation} shows the results of the ablation experiments. In order to deeply analyze the impact of different modules on report generation, we design several model variants for testing. Specifically, these include: \textbf{1) BASE:} The standard encoder-decoder model, consisting of a visual feature extractor and an encoder-decoder, is the basis for all variants.\textbf{ 2) TSG:} Based on the BASE model, a topic semantic guidance module is added to enhance semantic information extraction. \textbf{3) KSG{\scriptsize -B}:} Based on the BASE model, combined with keyword semantic guidance and BiomedCLIP is used to retrieve similar historical case reports. \textbf{4) TKSG{\scriptsize -B}:} It integrates both topic and keyword semantic guidance to enhance system performance. 

The results show that when the model uses only TSG, the model can understand the overall situation of the patient in advance compared to the base model, which leads to the generation of a more semantically correct report. And when using only KSG{\scriptsize-B}, the model is able to capture more detailed features in the image compared to the BASE model, making the generated content more precise and with keyword semantic relevance. The combination of these two in TKSG{\scriptsize-B} further enhances the overall performance of the model.

\begin{figure*}[t]
    \centering
    \includegraphics[width=\textwidth]{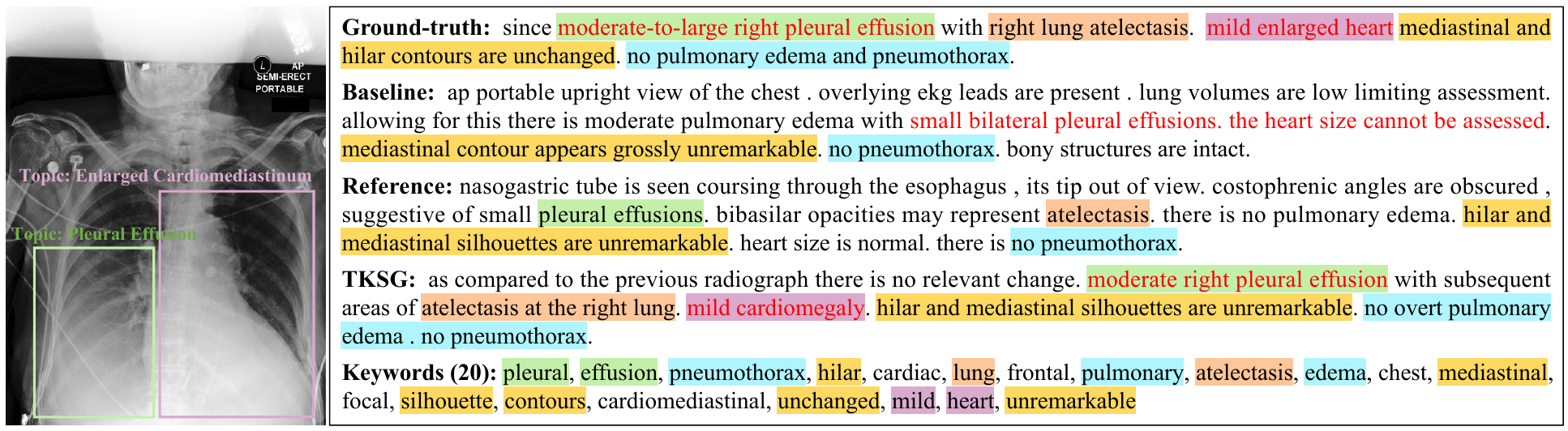} 
    \caption{Case study of TKSG. The same disease descriptions in the report and keywords are highlighted with the same background color. \textcolor[rgb]{1,0,0}{Red} text indicates the incorrect information generated by the baseline, while the content generated by TKSG is accurate.}
    \label{fig:case}
    \vspace{-15pt}
\end{figure*}

\subsection{Qualitative Analysis}

To investigate how our method predicts keywords and topic terms from visual and reference report information and aligns the two modalities, we conduct a case study shown in Figure~\ref{fig:case}.
In this case, fonts with the same color background indicate diagnoses that are consistent with the Ground Truth, red fonts indicate incorrect diagnoses generated by Baseline. TKSG succeeds in generating correct diagnoses. It correctly predicts all key descriptions in Ground-truth (e.g., \textit{the lungs are hyperinflated}, \textit{chronic obstructive pulmonary disease}, \textit{mild cardiomegaly} and \textit{right lung atelectasis}). In addition, compared to Baseline, TKSG also correctly recognizes Baseline's incorrect diagnosis (e.g., \textit{small bilateral pleural effusions} and \textit{the heart size cannot be assessed}). 
To illustrate how TKSG enhances diagnostic accuracy, we present its predicted topic terms and keywords. 
It can be seen that the topic vectors direct the model's attention towards the primary diseases, while the keywords offer essential information regarding disease symptoms. This combination effectively conveys diagnostic insights to the decoder throughout the generation process.
\section{Conclusion}
In this paper, we propose a novel radiological report generation method based on valid a priori information, namely the \textbf{T}opic-\textbf{K}eyword \textbf{S}emantic \textbf{G}uidance framework (TKSG). The framework consists of three core modules: the Topic Detector, the Keyword Detector and Semantic-Guided Attention mechanism. Among them, the Topic Detector and Keyword Detector are used to predict topic vectors and keywords, respectively.  As a guide for generating reports, the Semantic-Guided Attention mechanism realizes efficient alignment and interaction between text and images. Besides, we analyze the effects of hyper-parameter settings on model performance through a number of experiments. 
Experimental results on two benchmark datasets demonstrate the effectiveness of our model, which achieves the state-of-the-art performance.

\bibliographystyle{IEEEtran}
\bibliography{ref}

\begin{thebibliography}{10}
\providecommand{\url}[1]{#1}
\csname url@samestyle\endcsname
\providecommand{\newblock}{\relax}
\providecommand{\bibinfo}[2]{#2}
\providecommand{\BIBentrySTDinterwordspacing}{\spaceskip=0pt\relax}
\providecommand{\BIBentryALTinterwordstretchfactor}{4}
\providecommand{\BIBentryALTinterwordspacing}{\spaceskip=\fontdimen2\font plus
\BIBentryALTinterwordstretchfactor\fontdimen3\font minus \fontdimen4\font\relax}
\providecommand{\BIBforeignlanguage}[2]{{%
\expandafter\ifx\csname l@#1\endcsname\relax
\typeout{** WARNING: IEEEtran.bst: No hyphenation pattern has been}%
\typeout{** loaded for the language `#1'. Using the pattern for}%
\typeout{** the default language instead.}%
\else
\language=\csname l@#1\endcsname
\fi
#2}}
\providecommand{\BIBdecl}{\relax}
\BIBdecl

\bibitem{ganeshan2018structured}
D.~Ganeshan, P.-A.~T. Duong, L.~Probyn, L.~Lenchik, T.~A. McArthur, M.~Retrouvey, E.~H. Ghobadi, S.~L. Desouches, D.~Pastel, and I.~R. Francis, ``Structured reporting in radiology,'' \emph{Academic radiology}, vol.~25, no.~1, pp. 66--73, 2018.

\bibitem{zhang2023knowledge}
X.~Zhang, C.~Wu, Y.~Zhang, W.~Xie, and Y.~Wang, ``Knowledge-enhanced visual-language pre-training on chest radiology images,'' \emph{Nature Communications}, vol.~14, no.~1, p. 4542, 2023.

\bibitem{R2Gen}
Z.~Chen, Y.~Song, T.-H. Chang, and X.~Wan, ``Generating radiology reports via memory-driven transformer,'' \emph{arXiv preprint arXiv:2010.16056}, 2020.

\bibitem{R2GenCMN}
Z.~Chen, Y.~Shen, Y.~Song, and X.~Wan, ``Cross-modal memory networks for radiology report generation,'' \emph{arXiv preprint arXiv:2204.13258}, 2022.

\bibitem{M2KT}
S.~Yang, X.~Wu, S.~Ge, Z.~Zheng, S.~K. Zhou, and L.~Xiao, ``Radiology report generation with a learned knowledge base and multi-modal alignment,'' \emph{Medical Image Analysis}, vol.~86, p. 102798, 2023.

\bibitem{CheXbert}
A.~Smit, S.~Jain, P.~Rajpurkar, A.~Pareek, A.~Ng, and M.~Lungren, ``\BIBforeignlanguage{en-US}{Combining automatic labelers and expert annotations for accurate radiology report labeling using bert},'' in \emph{\BIBforeignlanguage{en-US}{Proceedings of the 2020 Conference on Empirical Methods in Natural Language Processing}}, Jan 2020.

\bibitem{biomedclip}
S.~Zhang, Y.~Xu, N.~Usuyama, H.~Xu, J.~Bagga, R.~Tinn, S.~Preston, R.~Rao, M.~Wei, N.~Valluri \emph{et~al.}, ``Biomedclip: a multimodal biomedical foundation model pretrained from fifteen million scientific image-text pairs,'' \emph{arXiv preprint arXiv:2303.00915}, 2023.

\bibitem{medclip}
Z.~Wang, Z.~Wu, D.~Agarwal, and J.~Sun, ``Medclip: Contrastive learning from unpaired medical images and text,'' \emph{arXiv preprint arXiv:2210.10163}, 2022.

\bibitem{clip}
A.~Radford, J.~W. Kim, C.~Hallacy, A.~Ramesh, G.~Goh, S.~Agarwal, G.~Sastry, A.~Askell, P.~Mishkin, J.~Clark \emph{et~al.}, ``Learning transferable visual models from natural language supervision,'' in \emph{International conference on machine learning}.\hskip 1em plus 0.5em minus 0.4em\relax PMLR, 2021, pp. 8748--8763.

\bibitem{IU}
D.~Demner-Fushman, M.~D. Kohli, M.~B. Rosenman, S.~E. Shooshan, L.~Rodriguez, S.~Antani, G.~R. Thoma, and C.~J. McDonald, ``Preparing a collection of radiology examinations for distribution and retrieval,'' \emph{Journal of the American Medical Informatics Association}, vol.~23, no.~2, pp. 304--310, 2016.

\bibitem{MIMIC}
A.~E. Johnson, T.~J. Pollard, N.~R. Greenbaum, M.~P. Lungren, C.-y. Deng, Y.~Peng, Z.~Lu, R.~G. Mark, S.~J. Berkowitz, and S.~Horng, ``Mimic-cxr-jpg, a large publicly available database of labeled chest radiographs,'' \emph{arXiv preprint arXiv:1901.07042}, 2019.

\bibitem{MultiCapCLIP}
B.~Yang, F.~Liu, X.~Wu, Y.~Wang, X.~Sun, and Y.~Zou, ``Multicapclip: Auto-encoding prompts for zero-shot multilingual visual captioning,'' \emph{arXiv preprint arXiv:2308.13218}, 2023.

\bibitem{miura2020improving}
Y.~Miura, Y.~Zhang, E.~B. Tsai, C.~P. Langlotz, and D.~Jurafsky, ``Improving factual completeness and consistency of image-to-text radiology report generation,'' \emph{arXiv preprint arXiv:2010.10042}, 2020.

\bibitem{kale2023replace}
K.~Kale, K.~Jadhav \emph{et~al.}, ``Replace and report: Nlp assisted radiology report generation,'' \emph{arXiv preprint arXiv:2306.17180}, 2023.

\bibitem{M2TR}
F.~Nooralahzadeh, N.~P. Gonzalez, T.~Frauenfelder, K.~Fujimoto, and M.~Krauthammer, ``Progressive transformer-based generation of radiology reports,'' \emph{arXiv preprint arXiv:2102.09777}, 2021.

\bibitem{FSE}
K.~Liu, Z.~Ma, M.~Liu, Z.~Jiao, X.~Kang, Q.~Miao, and K.~Xie, ``Factual serialization enhancement: A key innovation for chest x-ray report generation,'' \emph{arXiv preprint arXiv:2405.09586}, 2024.

\bibitem{swin}
Z.~Liu, H.~Hu, Y.~Lin, Z.~Yao, Z.~Xie, Y.~Wei, J.~Ning, Y.~Cao, Z.~Zhang, L.~Dong \emph{et~al.}, ``Swin transformer v2: Scaling up capacity and resolution,'' in \emph{Proceedings of the IEEE/CVF conference on computer vision and pattern recognition}, 2022, pp. 12\,009--12\,019.

\bibitem{CMM-RL}
H.~Qin and Y.~Song, ``Reinforced cross-modal alignment for radiology report generation,'' in \emph{Findings of the Association for Computational Linguistics: ACL 2022}, 2022, pp. 448--458.

\bibitem{MA}
H.~Shen, M.~Pei, J.~Liu, and Z.~Tian, ``Automatic radiology reports generation via memory alignment network,'' in \emph{Proceedings of the AAAI Conference on Artificial Intelligence}, vol.~38, no.~5, 2024, pp. 4776--4783.

\end{thebibliography}

\end{document}